\documentclass[prl,showpacs,floatfix,twocolumn]{revtex4-1}
\usepackage{amsfonts}
\usepackage{stmaryrd}
\usepackage{bbm}
\usepackage{mathrsfs}
\usepackage{tipa}
\usepackage{amssymb}
\usepackage{txfonts}
\usepackage{graphicx}
\usepackage{dcolumn}
\usepackage{epstopdf}
\usepackage[colorlinks,linkcolor=blue,urlcolor=blue,citecolor=blue]{hyperref}
\newcommand{\PreserveBackslash}[1]{\let\temp=\\#1\let\\=\temp}
\newcolumntype{C}[1]{>{\PreserveBackslash\centering}p{#1}}
\newcolumntype{R}[1]{>{\PreserveBackslash\raggedleft}p{#1}}
\newcolumntype{L}[1]{>{\PreserveBackslash\raggedright}p{#1}}

\begin{document}

\newcommand*{\cm}{cm$^{-1}$\,}

\title{Observation of anomalous temperature dependence of spectrum on small Fermi surfaces in a BiS$_{2}$-based superconductor}

\author{L. K. Zeng$^{1}$}
\author{X. B. Wang$^{1}$}
\author{J. Ma$^{1}$}
\author{P. Richard$^{1,2}$}
\author{S. M. Nie$^{1}$}
\author{H. M. Weng$^{1}$}
\author{N. L. Wang$^{1,2}$}
\author{Z. Wang$^{3}$}
\author{T. Qian$^{1}$}
\email{tqian@iphy.ac.cn}
\author{H. Ding$^{1,2}$}
\email{dingh@iphy.ac.cn}

\affiliation{$^1$Beijing National Laboratory for Condensed Matter Physics, and Institute of Physics, Chinese Academy of Sciences, Beijing 100190, China}
\affiliation{$^2$Collaborative Innovation Center of Quantum Matter, Beijing, China}
\affiliation{$^3$Department of Physics, Boston College, Chestnut Hill, Massachusetts 02467, USA}

%\date{\today}

\begin{abstract}

We have performed an angle-resolved photoemission spectroscopy study
of the BiS$_{2}$-based superconductor Nd(O,F)BiS$_{2}$. Two small
electron-like Fermi surfaces around X ($\pi$, 0) are observed, which
enclose 2.4$\%$ and 1.1$\%$ of the Brillouin zone area,
respectively, corresponding to an electron doping of 7$\%$ per Bi
site. The low-energy spectrum consists of a weakly-dispersing broad
hump and a dispersive branch, which follows well the calculated band
dispersion. This hump is drastically suppressed with increasing
temperature, while the dispersive branch is essentially unaffected.
The anomalous thermal effect indicates a highly interacting
electronic state, in which the superconducting pairing develops.

\end{abstract}


\maketitle

The recent discovery of superconductivity with \emph{T}$_{c}$ up to
$\sim$ 10 K in the BiS$_{2}$-based compounds has attracted a lot of
attentions \cite{M1,M2,M3,M4,M5,M6,M7,M8,M9,M10}. As in the cuprate
and iron-based high-\emph{T}$_{c}$ superconductors, the BiS$_{2}$
family has a layered crystal structure consisting of superconducting
BiS$_{2}$ layers intercalated with various block layers. Band
structure calculations show that the parent compound of the
BiS$_{2}$-based superconductors is a band insulator with an energy
gap of $\sim$ 0.8 eV \cite{C11,C12,C13,C14}, and bulk
superconductivity induced by electron doping is derived from the Bi
6\emph{p}$_{x}$/\emph{p}$_{y}$ orbitals, in which correlation
effects are expected to be weaker than those in the 3\emph{d}
orbitals of the cuprate and iron-based superconductors. The
superconducting transition temperature \emph{T}$_{c}$ reaches a
maximum at a nominal doping level $\delta$ $\sim$ 0.5 for many
compounds \cite{M1,M2,M3,M4,M5,M6,M7,M8,M9,M10}, where strong
nesting between the large parallel Fermi surface (FS) segments is
suggested in band calculations \cite{C11,C12,C13,C14}. Therefore,
most of the theoretical models for the pairing mechanism are based
on the nesting scenario.

However, there is a large bifurcation regarding the consequences of
the nesting. On the one hand, the nesting is proposed to enhance the
electron-phonon coupling, thus favoring a conventional BCS
superconductivity \cite{C11,C12,C13,C14}. On the other hand, as
widely believed for the iron-based superconductors, the strong FS
nesting could enhance charge or spin fluctuations, and thus
electronic correlations may play a major role in the superconducting
pairing \cite{C15,C16,C17,C18}. Magnetic penetration depth and
muon-spin rotation spectroscopy measurements support a conventional
\emph{s}-wave superconductivity in the strong electron-phonon
coupling limit \cite{P19,P20}, whereas the absence of phonon anomaly
in neutron scattering measurements suggests that the electron-phonon
coupling may be much weaker than theoretically expected \cite{P21}.
Recent scanning tunneling spectroscopy measurements show that the
ratio 2$\Delta$/\emph{k}$_{B}$\emph{T}$_{c}$ is much larger than the
BCS value \cite{P22,P23}. Giant superconducting fluctuations and an
anomalous semiconducting normal state are also observed, suggesting
that the superconductivity might be different from that of a
conventional BCS superconductor \cite{P23}.

In this Letter, we present angle-resolved photoemission spectroscopy
(ARPES) results of the BiS$_{2}$-based superconductor
Nd(O,F)BiS$_{2}$ (\emph{T}$_{c}$$^{zero}$ = 4 K). Two small
electron-like FSs around X ($\pi$, 0) are observed, corresponding to
an electron doping of 7$\%$ of itinerant carriers per Bi site. As a
result, the measured electronic structure is far from the proposed
FS nesting. Furthermore, we reveal that the low-energy spectrum
consists of a broad hump around -0.3 eV, which is drastically
suppressed with increasing temperature, and a dispersive branch,
which is essentially unaffected by the temperature. This exotic
spectral behavior suggests that the low-temperature normal state of
this superconductor is a highly interacting electronic state.

Single crystals with a nominal composition of
NdO$_{0.7}$F$_{0.3}$BiS$_{2}$ were grown by a flux method with
KCl/LiCl as the flux. Energy dispersion spectrum (EDS) measurements
were performed on several pieces of samples, which give an averaged
composition of
Nd$_{0.95\pm0.02}$O$_{y}$F$_{0.44\pm0.1}$Bi$_{0.94\pm0.02}$S$_{2}$.
ARPES measurements were performed at the Institute of Physics,
Chinese Academy of Sciences, using the He I$\alpha$ (\emph{h}$\nu$ =
21.218 eV) resonance lines. The angular and energy resolutions were
set to 0.2$^{\circ}$ and 14 $\sim$ 32 meV, respectively. Samples
with a typical size of $\sim$ 1 $\times$ 1 mm$^{2}$ were cleaved
\emph{in situ} at 30 K and measured between 30 and 230 K in a
working vacuum better than 4 $\times$ 10$^{-11}$ Torr. The Fermi
level (\emph{E}$_{F}$) of the samples was referenced to that of a
gold film evaporated onto the sample holder.

\begin{figure}
\includegraphics[width=3.3in]{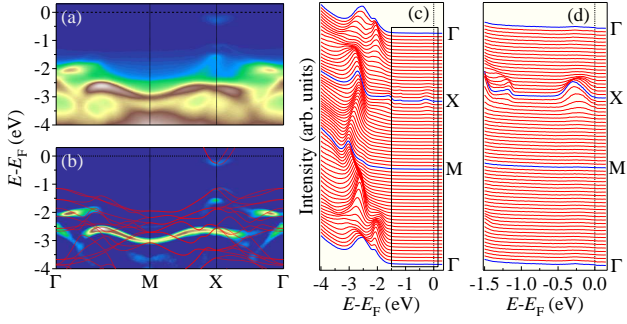}
\caption{(Color online) (a) ARPES intensity plot along the high
symmetry lines $\Gamma$-M-X-$\Gamma$ taken at 30 K. $\Gamma$-X is
along the nearest Bi-Bi direction. (b) Corresponding two-dimensional
curvature intensity plot. LDA bands without SOC for the undoped
parent compound LaOBiS$_{2}$, which are essentially the same as
those of NdOBiS$_{2}$ in the energy in our experiments, are also
plotted in (b) for comparison. \emph{E}$_{F}$ of the calculated band
structure is set to be 1.15 eV above the top of the valence bands to
match the experimental band dispersions. (c) Corresponding EDCs. (d)
Magnification of the box in (c).}
\end{figure}

Figure 1 shows the band dispersions along the high-symmetry lines
$\Gamma$-M-X-$\Gamma$ in an energy range within 4 eV below
\emph{E}$_{F}$. We observe several dispersive bands below -1.2 eV
and an electron-like band dispersion with a bottom of -0.3 eV near
X. There is an energy gap of $\sim$ 0.9 eV between them. To
understand the multiband electronic structure, we superimpose the
local-density approximation (LDA) band structure on top of our data.
The calculated band structure reflects some main features in the
experiment data, especially for the direct gap between the
conduction and valence bands. In LDA calculations, the undoped
parent compound is a band insulator and its \emph{E}$_{F}$ is
located within the energy gap. The experimentally obtained
\emph{E}$_{F}$ is situated in the conduction bands, indicating that
electron carriers are introduced in the superconducting samples due
to the substitution of O with F.

Figure 2 shows the FS mapping data in the
\emph{k}$_{x}$-\emph{k}$_{y}$ plane. We extract two FS pockets
centered at X, which come from the near-\emph{E}$_{F}$ electron-like
dispersion shown in Fig. 1. The two extracted FSs exhibit a
significantly anisotropic separation, which is a result of the
cooperative effects from spin-orbit coupling (SOC) and interlayer
coupling, as explained below. We have performed LDA calculations
using the following models. In a one-BiS$_{2}$-layer model (same as
the surface layer after cleave) with SOC [Fig. 2(d)], the
near-\emph{E}$_{F}$ bands are split along both XM and $\Gamma$X with
comparable magnitudes. In a two-BiS$_{2}$-layer model (same as the
bulk) without SOC [Fig. 2(e)], the bands are doubly-degenerate along
XM, but split along $\Gamma$X due to the interlayer coupling. In the
presence of SOC, the Rashba term lifts the degeneracy along XM but
the splitting magnitude along XM is much smaller than along
$\Gamma$X [Fig. 2(f)], in agreement with our observations.

\begin{figure}
\includegraphics[width=3.3in]{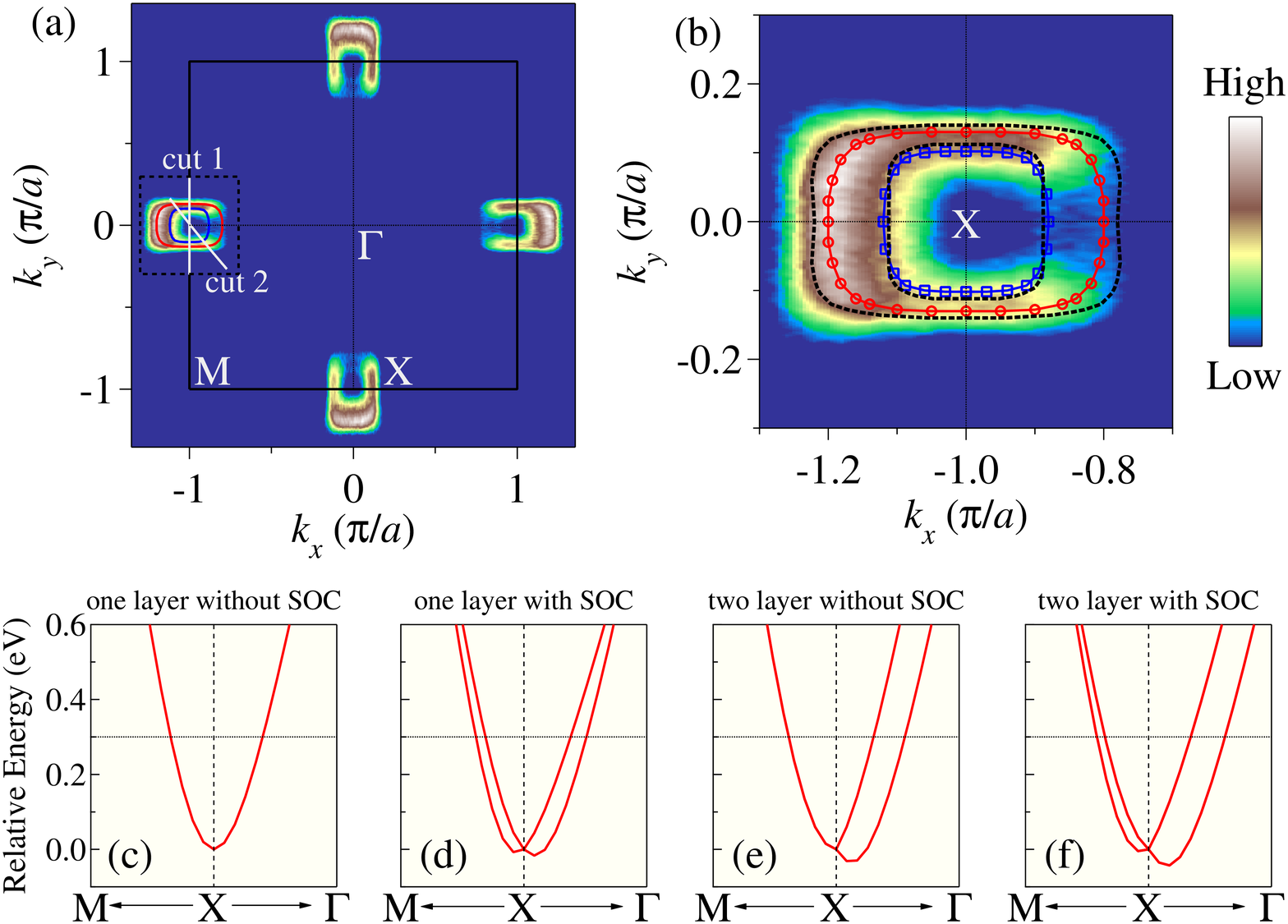}
\caption{(Color online) (a) ARPES intensity plot at \emph{E}$_{F}$
as a function of the two-dimensional wave vector taken at 30 K. The
intensity is obtained by integrating the spectra within $\pm$15 meV
with respect to \emph{E}$_{F}$. The data are symmetrized by assuming
a fourfold symmetry with respect to $\Gamma$. Red and blue lines
represent the extracted FSs. White lines labeled cut1 and cut2
indicate the momentum locations, along which the data are shown in
Figs. 3 and 4, respectively. (b) Magnification of the dashed box in
(a). Circles and squares show the experimentally determined
\emph{k}$_{F}$ points. Dashed black lines represent the LDA + SOC
calculated FSs. (c) and (d) Calculated conduction bands at X using
one-BiS$_{2}$-layer models with and without SOC, respectively. (e)
and (f) Same as (c) and (d), respectively, but using
two-BiS$_{2}$-layer models. In (c)-(f), the zero energy is defined
to the band bottom at X and the horizontal dashed lines at 0.3 eV
represent the approximate position of \emph{E}$_{F}$.}
\end{figure}

The two FS pockets enclose 1.1$\%$ and 2.4$\%$ of the Brillouin zone
area, respectively. Counting the Luttinger volume of two-dimensional
FS sheets, the two observed FSs correspond to an electron doping of
7$\%$ per Bi site. The value for doped itinerant carrier density is
much less than those inferred from the nominal composition and the
EDS data. The discrepancy can be explained in several ways. Firstly,
the possibility of charge polarization at the terminal layer cannot
be completely excluded, though this scenario is unlikely since the
cleavage occurs between two symmetrical BiS$_{2}$ layers. Moreover,
the plasma frequency calculated using the experimental doping level
is $\sim$ 2.1 eV, in agreement with the optical data \cite{P24},
suggesting that the ARPES data reflect the intrinsic carrier density
in the bulk. Secondly, as both oxygen and fluorine are light
elements, their concentrations given from the EDS data may not be
reliable \cite{P23}. Thirdly, part of the carriers may be localized
and thus do not contribute to the conduction band. In this case, the
localized carriers could form flat bands within the energy gap.

\begin{figure}
\includegraphics[width=3.3in]{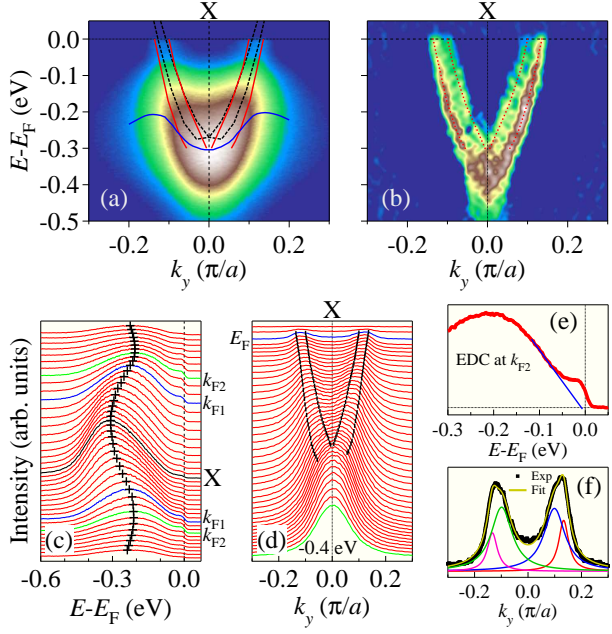}
\caption{(Color online) (a) ARPES intensity plot along XM (cut 1
from Fig. 2(a)) taken at 30 K. Red and blue lines represent the
dispersions extracted from the peak positions of the MDCs and EDCs,
respectively. Dashed black lines represent the LDA + SOC calculated
bands. (b) Corresponding intensity plot of second derivative along
momentum. Dotted lines are the same as the red lines in (a). (c)
Corresponding EDCs. Crosses indicate the peak positions of the EDCs.
Black, blue and green curves represent the EDCs at X,
\emph{k}$_{F1}$ and \emph{k}$_{F2}$, respectively. \emph{k}$_{F1}$
and \emph{k}$_{F2}$ represent the Fermi wave vectors of the inner
and outer electron-like bands, respectively. (d) Corresponding MDCs.
Short verticals indicate the peak positions of the MDCs. Blue and
green curves represent the MDCs at \emph{E}$_{F}$ and -0.4 eV,
respectively. (e) EDC at \emph{k}$_{F2}$. Blue line represents a
linear extrapolation of the slope on the lower binding energy side
of the hump. (f) MDC at \emph{E}$_{F}$. The MDC is fitted to four
Lorentzian peaks, indicating the band splitting along XM due to
SOC.}
\end{figure}

We find that at low temperature, the low-energy spectrum near
\emph{E}$_{F}$ consists of a large hump around -0.3 eV and a
dispersive branch, which tracks well the calculated conduction band
dispersion. Figure 3 shows ARPES data of the conduction bands taken
along XM at 30 K. Two electron-like bands are resolved in Figs. 3(b)
and 3(d). As mentioned above, the band splitting along XM originates
from the SOC. The band dispersion, extracted by tracking the peak
positions of the momentum distribution curves (MDCs), follows well
the LDA bands calculated with SOC. On the other hand, the energy
distribution curves (EDCs) are characterized by a broad hump, whose
maximum does not cross \emph{E}$_{F}$ but tends to bend back beyond
\emph{k}$_{F2}$ [Fig. 3(c)]. As shown in Fig. 3(e), the EDC at
\emph{k}$_{F2}$ shows a sharp Fermi cutoff, indicating a metallic
behavior. There is a change of the slope on the lower binding energy
side of the hump at $\sim$ -0.05 eV. The linear extrapolation
suggests that the hump contributes vanishingly small spectral weight
at \emph{E}$_{F}$, indicating that the finite low-energy spectral
weight at \emph{E}$_{F}$ is dominated by the dispersive branch.

\begin{figure}
\includegraphics[width=3.3in]{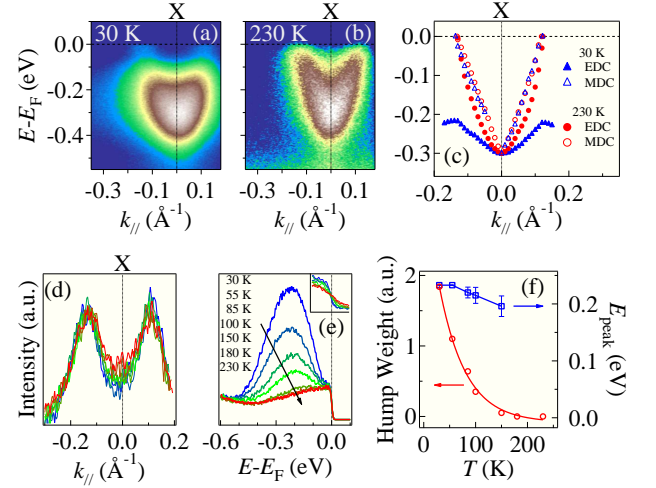}
\caption{(Color online) (a) and (b) ARPES intensity plots through
the X point (cut 2 from Fig. 2(a)) taken at 30 and 230 K,
respectively. (c) \emph{E}-\emph{k} plot of the positions of the EDC
and MDC peaks taken at 30 and 230 K, respectively. To approximately
remove the effect of the Fermi function on the EDC peak position
near \emph{E}$_{F}$, the EDC peak positions are extracted from the
symmetrized curve with respect to \emph{E}$_{F}$. (d) and (e) MDCs
at \emph{E}$_{F}$ and EDCs at \emph{k}$_{F}$ of the left branch
taken at various temperatures between 30 and 230 K, respectively.
Thermal broaden effect on the EDCs in (e) is removed (see text for
details). Inset of (e) plots the EDCs (raw data) in a energy window
of [-0.05, 0.05 eV]. (f) Spectral weight (left axis) and binding
energy of the maximum (right axis) of the broad hump against
temperature. The spectral weight is obtained by subtracting the
integration of the EDC at 230 K from that taken at the corresponding
temperature.}
\end{figure}

The low-energy spectrum shows anomalous temperature dependence
characterized by a rapid suppression of the spectral weight of the
broad hump with increasing temperature that, nevertheless, leaves
the dispersive branch little changed. The temperature dependent
ARPES results along cut 2 are shown in Fig. 4. A sharp contrast
between the intensity contours at 30 K [Fig. 4(a)] and 230 K [Fig.
4(b)] is clearly visible. To further clarify the evolution of the
spectrum with temperature, we plot the MDCs at \emph{E}$_{F}$ and
the EDCs at \emph{k}$_{F}$ of the left branch at various
temperatures between 30 and 230 K in Figs. 4(d) and 4(e),
respectively. All the spectra are normalized by the photon flux. To
remove the thermal broadening effect due to the Fermi-Dirac
statistics, the EDCs in Fig. 4(e) are divided by the
resolution-convoluted Fermi functions at corresponding temperatures
and then multiplied by that of 10 K. The MDCs almost collapse onto a
single curve, whereas the lineshape of the EDCs changes dramatically
with temperature. The spectral weight of the broad hump is
suppressed rapidly with increasing temperature. The hump can no
longer be clearly resolved above 150 K and the spectra at 180 K and
230 K nearly coincide. In sharp contrast to the loss of spectral
weight below \emph{E}$_{F}$, the Fermi cutoff of all the EDCs (raw
data) taken at various temperatures crosses exactly at
\emph{E}$_{F}$ [inset of Fig. 4(e)], which is consistent with the
collapse of the MDCs shown in Fig. 4(d), indicating a negligible
temperature effect on the spectral weight at \emph{E}$_{F}$ that is
dominated by the dispersive branch. As shown in Fig. 4(c), with the
suppression of the broad hump, the dichotomy between the dispersions
of EDC and MDC peaks is almost eliminated at 230 K. The dispersions
at 230 K are in good consistence with the one extracted from the
MDCs at 30 K that corresponds to the dispersive branch. Therefore,
in sharp contrast to the strong temperature dependence of the broad
hump, the dispersive branch is essentially unaffected with
temperature.

A conventional explanation for temperature-induced loss of spectral
weight is through the effect of lattice vibrations. Such an effect
in photoemission spectra is similar to that found in X-ray and
neutron scattering, where the intensities of diffraction peaks are
multiplied by the Debye-Waller factor \emph{e}$^{-2W}$ (\emph{W}
$\propto$ \emph{T}) \cite{E25}. Indeed, the integrated spectral
weight as a function of temperature can be approximately fitted to a
function of \emph{A} + \emph{Be}$^{-CT}$, as shown in Fig. 4(f).
However, the extracted Debye temperature from the fitting is about
only 3 K \cite{E26}, two orders of magnitudes smaller than the
estimated values from the specific heat data \cite{M6,M7,D1,D2,D3}.
Therefore, the effect of conventional lattice vibrations cannot
explain the giant temperature effects in our data.

Loss of spectral weight over a large energy scale has been observed
in a polaronic state of the colossal magnetoresistant manganites
La$_{2-2x}$Sr$_{1+2x}$Mn$_{2}$O$_{7}$ \cite{E27,E28,E29,E30}. In the
polaronic state induced by strong electron-phonon coupling, the
spectral function consists of a low-energy ``zero-phonon"
quasiparticle peak and a hump-like high-energy incoherent resonance.
Our observation of a high-energy hump as well as a sharp Fermi
cutoff bears some resemblance to the signature of polarons. In the
manganites, the incoherent branch loses its partial spectral weight
over an energy range of up to 0.8 eV, which is accompanied by a
disappearance of the quasiparticle peaks around the metal-insulator
transition temperature. This could be associated with either the
loss of polaron coherence \cite{E29} or a decreased fraction of
metallic regions in the scenario of phase separation \cite{E28}.
Assuming that the polaron picture could be applied to the BiS$_{2}$
system, the spectral weight of the incoherent branch, \emph{i.e.}
the broad hump, is drastically suppressed with increasing
temperature, while the low-energy coherent branch, which dominates
the spectral weight at \emph{E}$_{F}$, is not affected. The behavior
looks very unusual in the framework of polarons because the strong
suppression of the incoherent branch indicates significant changes
of the polaronic states, which seems not to be perceived by the
coherent part. Our spectra exhibit distinctly different temperature
dependence from those in the manganites, indicating that the giant
thermal effects observed in the two systems might have different
origins.

The disorder-induced self-traping of polarons in
Na$_{0.025}$WO$_{3}$ also shows nontrivial temperature dependence of
the spectral function \cite{E31,E32}. The Na$_{x}$WO$_{3}$ and
BiS$_{2}$ systems bear an interesting resemblance in their
electronic structures. Their undoped parent compounds are band
insulators with energy gaps of an order of eV between the valence
and conduction bands. The introduction of electron carriers by
element substitutions or intercalations leads to an insulator-metal
transition on small electron-like FSs. The conduction electrons in
Na$_{0.025}$WO$_{3}$ are self-trapped due to strong disorder induced
by the randomly distributed Na$^{+}$ ions, forming a weakly
dispersive polaron band near the top of valence bands. The breakdown
of polarons at high temperature leads to a large decrease in the
intensity of the polaron band, while the spectrum of the conduction
band is not significantly changed. These properties are quite
similar to what we have observed in BiS$_{2}$ except that the
polaron band would be localized near the bottom of the conduction
bands in our case. We note that the polaronic self-trapping of
carriers could also explain the small FS pockets observed in our
experiment. It is expected that as the polarons become more
delocalized, the spectral weight should be transferred to other
\emph{k}-points. However, such a spectral weight transfer is not
observed at least along the measured momentum cut.

In summary, our ARPES results show two small electron-like FSs
around X ($\pi$, 0) instead of large hole-like FSs centered at
$\Gamma$ (0, 0) and M ($\pi$, $\pi$) proposed in recent theoretical
models. The anomalous temperature dependence of the low-energy
spectrum indicates that the superconducting pairing develops in a
highly interacting electronic state. Our results provide detailed
information on the low-energy electronic structures and valuable
insights for further experimental and theoretical studies of the
pairing mechanism in the BiS$_{2}$-based superconductors.

We acknowledge X. Dai and Z. Fang for valuable discussions. This
work was supported by grants from CAS (2010Y1JB6), MOST
(2010CB923000, 2011CBA001000, 2011CB921701, 2013CB921700,
2011CBA00108, and 2012CB821403), NSFC (11004232, 11050110422,
11274362, 11234014, 11120101003, 11074291, 11274359, and 11104339),
and DOE (DE-FG02-99ER45747 and DE-SC0002554).

\bibliographystyle{apsrev4-1}

\begin{references}
\bibitem{M1} Y. Mizuguchi, H. Fujihisa, Y. Gotoh, K. Suzuki, H. Usui, K. Kuroki, S. Demura, Y. Takano, H. Izawa, and O. Miura, Phys. Rev. B \textbf{86}, 220510(R) (2012).
\bibitem{M2} S. K. Singh, A. Kumar, B. Gahtori, Shruti, G. Sharma, S. Patnaik, and V. P. S. Awana, J. Am. Chem. Soc. \textbf{134}, 16504 (2012).
\bibitem{M3} Y. Mizuguchi, S. Demura, K. Deguchi, Y. Takano, H. Fujihisa, Y. Gotoh, H. Izawa, and O. Miura, J. Phys. Soc. Jpn. \textbf{81}, 114725 (2012).
\bibitem{M4} S. Demura, Y. Mizuguchi, K. Deguchi, H. Okazaki, H. Hara, T. Watanabe, S. J. Denholme, M. Fujioka, T. Ozaki, H. Fujihisa, Y. Gotoh, O. Miura, T. Yamaguchi, H. Takeya, and Y. Takano, J. Phys. Soc. Jpn. \textbf{82}, 033708 (2013).
\bibitem{M5} J. Xing, S. Li, X. Ding, H. Yang, and H. H. Wen, Phys. Rev. B \textbf{86}, 214518 (2012).
\bibitem{M6} X. Lin, X. X. Ni, B. Chen, X. F. Xu, X. X. Yang, J. H. Dai, Y. K. Li, X. J. Yang, Y. K. Luo, Q. Tao, G. H. Cao, and Z. A. Xu, Phys. Rev. B \textbf{87}, 020504(R) (2013).
\bibitem{M7} D. Yazici, K. Huang, B. D. White, I. Jeon, V. W. Burnett, A. J. Friedman, I. K. Lum, M. Nallaiyan, S. Spagna, and M. B. Maple, Phys. Rev. B \textbf{87}, 174512 (2013).
\bibitem{M8} K. Deguchi, Y. Mizuguchi, S. Demura, H. Hara, T. Watanabe, S. J. Denholme, M. Fujioka, H. Okazaki, T. Ozaki, H. Takeya, T. Yamaguchi, O. Miura, and Y. Takano, Europhys. Lett. \textbf{101}, 17004 (2013).
\bibitem{M9} S. Demura, K. Deguchi, Y. Mizuguchi, K. Sato, R. Honjyo, A. Yamashita, T. Yamaki, H. Hara, T. Watanabe, S. J. Denholme, M. Fujioka, H. Okazaki, T. Ozaki, O. Miura, T. Yamaguchi, H. Takeya, and Y. Takano, arXiv:1311.4267
\bibitem{M10} R. Jha and V. P. S. Awana, arXiv:1401.4811
\bibitem{C11} H. Usui, K. Suzuki, and K. Kuroki, Phys. Rev. B \textbf{86}, 220501(R) (2012).
\bibitem{C12} X. G. Wan, H. C. Ding, S. Y. Savrasov, and C. G. Duan, Phys. Rev. B \textbf{87}, 115124 (2013).
\bibitem{C13} B. Li, Z. W. Xing, and G. Q. Huang, Europhys. Lett. \textbf{101}, 47002 (2013).
\bibitem{C14} T. Yildirim, Phys. Rev. B \textbf{87}, 020506(R) (2013).
\bibitem{C15} T. Zhou and Z. D. Wang, J. Supercond. Novel Magn. \textbf{26}, 2735 (2013).
\bibitem{C16} G. B. Martins, A. Moreo, and E. Dagotto, Phys. Rev. B \textbf{87}, 081102(R) (2013).
\bibitem{C17} Y. Yang, W. S. Wang, Y. Y. Xiang, Z. Z. Li, and Q. H. Wang, Phys. Rev. B \textbf{88}, 094519 (2013).
\bibitem{C18} Y. Liang, X. X. Wu, W. F. Tsai, and J. P. Hu, arXiv:1211.5435
\bibitem{P19} G. Lamura, T. Shiroka, P. Bonfa, S. Sanna, R. De Renzi, C. Baines, H. Luetkens, J. Kajitani, Y. Mizuguchi, O. Miura, K. Deguchi, S. Demura, Y. Takano, and M. Putti, Phys. Rev. B \textbf{88}, 180509(R) (2013).
\bibitem{P20} Shruti, P. Srivastava, and S. Patnaik, J. Phys.: Condens. Matter \textbf{25}, 339601 (2013).
\bibitem{P21} J. Lee, M. B. Stone, A. Huq, T. Yildirim, G. Ehlers, Y. Mizuguchi, O. Miura, Y. Takano, K. Deguchi, S. Demura, and S. H. Lee, Phys. Rev. B \textbf{87}, 205134 (2013).
\bibitem{P22} S. Li, H. Yang, D. Fang, Z. Wang, J. Tao, X. Ding, and H. H. Wen, Sci. China-Phys. Mech. Astron. \textbf{56}, 2019 (2013).
\bibitem{P23} J. Z. Liu, D. L. Fang, Z. Y. Wang, J. Xing, Z. Y. Du, X. Y. Zhu, H. Yang, and H. H. Wen, arXiv:1310.0377
\bibitem{P24} X. B. Wang and N. L. Wang (unpublished)
\bibitem{E25} S. H\"{u}fner, \emph{Photoelectron Spectroscopy: Principles and Application} (Springer-Verlag, New York, 1995).
\bibitem{E26} In the calculation, $\Delta$\emph{k} is defined to be (0, 0, 2$\pi$/\emph{c}).
\bibitem{D1} H. Takatsu, Y. Mizuguchi, H. Izawa, O. Miura, and H.
Kadowaki, J. Phys. Soc. Jpn. \textbf{81}, 125002 (2012).
\bibitem{D2} R. Jha, A. Kumar, S. K. Singh, and V. P. S. Awana, J.
Appl. Phys. \textbf{113}, 056102 (2013).
\bibitem{D3} D. Yazici, K. Huang, B. D. White, A. H. Chang, A. J.
Friedman, M. B. Maple, Philos. Mag. \textbf{93}, 673 (2013).
\bibitem{E27} S. de Jong, Y. Huang, I. Santoso, F. Massee, R. Follath, O. Schwarzkopf, L. Patthey, M. Shi, and M. S. Golden, Phys. Rev. B \textbf{76}, 235117 (2007).
\bibitem{E28} Z. Sun, J. F. Douglas, A. V. Fedorov, Y. D. Chuang, H. Zheng, J. F. Mitchell, and D. S. Dessau, Nature Phys. \textbf{3}, 248 (2007).
\bibitem{E29} N. Mannella, W. L. Yang, X. J. Zhou, H. Zheng, J. F. Mitchell, J. Zaanen, T. P. Devereaus, N. Nagaosa, Z. Hussain, and Z. X. Shen, Nature \textbf{438}, 474 (2005).
\bibitem{E30} N. Mannella, W. L. Yang, K. Tanaka, X. J. Zhou, H. Zheng, J. F. Mitchell, J. Zaanen, T. P. Devereaux, N. Nagaosa, Z. Hussain, and Z. X. Shen, Phys. Rev. B \textbf{76}, 233102 (2007).
\bibitem{E31} S. Raj, D. Hashimoto, H. Matsui, S. Souma, T. Sato, T. Takahashi, D. D. Sarma, P. Mahadevan, and S. Oishi, Phys. Rev. Lett. \textbf{96}, 147603 (2006).
\bibitem{E32} S. Raj, T. Sato, S. Souma, and T. Takahashi, Mod. Phys. Lett. B \textbf{23}, 2819 (2009).
\end{references}

\end{document}